\documentclass{elsart}
\usepackage{amsmath}
\usepackage{epsfig}
\usepackage{hyperref}

\usepackage{graphicx}
\usepackage[usenames]{color}
\usepackage[normalem]{ulem}

\begin{document}
\begin{frontmatter}

\title{Manipulation of light in a generalized coupled Nonlinear
Schr\"{o}dinger equation}

\author{R.Radha$^{\ast,1}$}
\corauth{Corresponding author.}
 \ead{radha\_ramaswamy@yahoo.com, \\
 Telephone: (91)-0435-2403119, Fax: (91)-0435-2403119}
\author{P.S.Vinayagam$^1$}
\address{$^1$ Centre for Nonlinear Science (CeNSc), PG and Research Department of Physics,
Government College for Women (Autonomous), Kumbakonam 612001,
India}

\author{K. Porsezian$^{\dagger,2}$}
\ead{ponzsol@yahoo.com$^{\dagger }$}
\address{$^2$ Department of Physics, Pondicherry University,
Puducherry-605014, India.}

\begin{abstract}
We investigate a generalized coupled nonlinear Schrodinger (GCNLS)
equation containing Self-Phase Modulation (SPM), Cross-Phase
Modulation (XPM) and Four Wave Mixing (FWM) describing the
propagation of electromagnetic radiation through an optical fibre
and generate the associated Lax-pair. We then construct bright
solitons employing gauge transformation approach. The collisional
dynamics of bright solitons indicates that it is not only possible
to manipulate intensity (energy) between the two modes (optical
beams), but also within a given mode unlike the Manakov model
which does not have the same freedom. The freedom to manipulate
intensity (energy) in a given mode or between two modes arises due
to a suitable combination of SPM, XPM and FWM. While SPM and XPM
are controlled by an arbitrary real parameter each, FWM is
governed by two arbitrary complex parameters. The above model may
have wider ramifications in nonlinear optics and Bose-Einstein
Condensates (BECs).
\end{abstract}
\end{frontmatter}
\newpage

\section{Introduction}
The potential of solitons to carry information through optical
fibres governed by the nonlinear Schrodinger (NLS) equation
\cite{nlseqn,nlseqn2,fibre,fibre2} and the freedom to switch
energy between two laser beams in a fibre described by the
celebrated Manakov model has made a dramatic turnaround in the
field of optical communications. The concept of shape changing
collisional dynamics of solitons in the Manakov model is governed
by the coupled NLS equation of the following form
\begin{subequations}
\begin{align}
i q_{1t}+q_{1xx}+2(g_{11}|q_1|^2
+g_{12}|q_2|^2)q_1 =0, \\
i q_{2t}+q_{2xx}+2(g_{21}|q_1|^2 + g_{22} |q_2|^2)q_2 =0,
\end{align}\label{nlsmode1}
\end{subequations}
where $q_{i}(x,t)$(i=1,2) corresponds to the envelope of the
electromagnetic radiation passing through the optical fibre and
$i$ is the imaginary unit. In the above equation, $g_{11}$ and
$g_{22}$ correspond to the Self Phase Modulation (SPM) while
$g_{12}$ and $g_{21}$ represent the Cross Phase Modulation (XPM).
Equations.(\ref{nlsmode1}) have been shown to be integrable and
admit painlev\'{e} property \cite{painleve} if either (i)
$g_{11}=g_{12}=g_{21}=g_{22}$ or (ii)
$g_{11}=g_{21}=-g_{12}=-g_{22}$. The first choice corresponds to
the celebrated Manakov model \cite{manakov,manakov2}  while the
second choice represents the modified Manakov model
\cite{modifymanakov,modifymanakov2} and it has been observed that
both admit shape changing collisional dynamics of bright solitons
\cite{rkrishnan}. Recently, it was shown \cite{psv:pre:nls} that
one can rotate the trajectories of the bright solitons by varying
the system parameters, namely SPM and XPM without violating the
integrability of the Manakov (or modified Manakov) model.

It should be mentioned that the inelastic collision of bright
solitons which is concerned with redistribution of energy between
two modes (optical beams) is brought about by varying the
parameters associated with the phase of bright solitons combined
with a suitable combination of coupling coefficients. Can one
manipulate the intensity (or energy) in a given mode
(electromagnetic radiation) or in a given bound state of the
electromagnetic radiation $?$. Can one manipulate optical pulses
by varying its interaction with the medium rather than changing
the parameters associated with  the phase of solitons $?$. The
answer to these questions assumes tremendous significance as one
will have the flexibility of desirably energizing a given mode  or
a given bound state of the optical pulse as two laser beams
propagate through optical fibres. This situation is reminescent of
manipulating binary interaction through Feshbach resonance
\cite{FR} in Bose-Einstein Condensates \cite{review}.

In addition to Manakov (or modified Manakov) model, a generalized
coupled NLS equation (GCNLS) by including Four Wave Mixing (FWM)
with SPM and XPM has been investigated by Park and Shin
\cite{park.shin} which subsequently led to the identification of
four different classes of integrable models dealing with the
propagation of optical beams through birefringent fibres.
Eventhough the variants of the GCNLS equation were investigated
recently by Wang and Agalar et al.,\cite{motopaper,agalarov}
respectively, with three arbitrary parameters (two real parameters
corresponding to SPM and XPM and one complex parameter for FWM),
the impact of FWM when reinforced with SPM and XPM on the
collisional dynamics of solitons  has not yet been clearly spelt
out. In addition, the impact of the freedom associated with four
arbitrary parameters in a GCNLS has not been probed yet.

In this paper, we investigate a GCNLS equation involving four
system parameters. The two real arbitrary parameters are
associated with SPM and XPM while the two arbitrary complex
parameters  are associated with Four Wave Mixing (FWM). We then
construct the Lax-pair of the GCNLS equation and generate bright
solitons. We then show that one can not only manipulate the
intensity (or energy)  between two laser beams, but also manoeuvre
the energy distribution among the bound states of a given laser
beam. The freedom to manipulate intensity arises from a suitable
combination of nonlinear interaction parameters associated with
the system.

\section{Mathematical model and Lax-Pair}
We know that the coupling between co-propagating optical beams in
a nonlinear medium determines the application of optical fibres.
Considering the propagation of optical pulses through a nonlinear
birefringent fibre, the dynamics is governed by the generalized
coupled NLS (GCNLS) equation of the following form
\begin{subequations}
\begin{align}
i \psi_{1t}+\psi_{1xx}+2(a|\psi_1|^2+c|\psi_2|^2+b\psi_1\psi_2^{\ast}+d\psi_2\psi_1^*)\psi_1=0, \\
i
\psi_{2t}+\psi_{2xx}+2(a|\psi_1|^2+c|\psi_2|^2+b\psi_1\psi_2^{\ast}+d\psi_2\psi_1^*)\psi_2=0,
\end{align}\label{gptwo}
\end{subequations}
In equations.(\ref{gptwo}), $\psi_1$ and $\psi_2$ represent the
strengths of electromagnetic beams. The nonlinear co-efficients
$a$ and $c$ which are real account for the SPM and XPM
respectively while the arbitrary real parameters $b$ and $d$
correspond to FWM. It should be mentioned that eventhough one can
allow FWM parameters $b$ and $d$  to be complex, we have retained
them to be real in the present model. When FWM effects ($b$ and
$d$) are equal to zero and the co-efficients $a = c$, then the
above model reduces to the celebrated Manakov or modified Manakov
\cite{manakov,modifymanakov} model. Equation.(\ref{gptwo}) has
also been investigated recently for $d= b^*$ \cite{motopaper} and
the impact of FWM on the collisional dynamics of solitons has been
investigated.  The above equation (\ref{gptwo}) admits the
following linear eigenvalue problem of the following form,

\begin{align}
\Phi_x &+ {\cal U} \Phi=0,\\
\Phi_t &+ {\cal V} \Phi=0,
\end{align}\label{lax}
where $\Phi = (\phi_1, \phi_2, \phi_3)^T$ and
\begin{align}
{\cal U} &= \left(%
\begin{array}{ccc}
i\zeta_1 & \psi_{1} &  \psi_{2}\\
-R_{1}& -i\zeta_1 & 0 \\
-R_{2}& 0 & -i\zeta_1 \\
\end{array}%
\right),
\end{align}

\begin{align}
{\cal V}&=\left(%
\begin{array}{ccc}
-i \zeta_1^{2}+ \frac{i}{2} \psi_{1} R_{1}+ \frac{i}{2} \psi_{2}R_{2} & - \zeta_1 \psi_{1}+ \frac{i}{2} \psi_{1x} & - \zeta_1 \psi_{2}+ \frac{i}{2} \psi_{2x} \\
\zeta_{1} R_{1} + \frac{i}{2} R_{1x} & i \zeta_{1}^{2} - \frac{i}{2} \psi_{1} R_{1} & -\frac{i}{2} \psi_{2} R_{1} \\
\zeta_{1} R_{2} + \frac{i}{2} R_{2x} & -\frac{i}{2} \psi_{1} R_{2} & i \zeta_{1}^{2} - \frac{i}{2} \psi_{2} R_{2} \\
\end{array}%
\right),
\end{align}
and
\begin{align}
R_{1}&= -a \psi_1(x,t)^{\ast}-b \psi_2(x,t)^{\ast}, \notag\\
R_{2}&= -d \psi_1(x,t)^{\ast}-c \psi_2(x,t)^{\ast}. \notag
\end{align}

In the above equation, the spectral parameter $\zeta_1$ is
isospectral. It is obvious that the compatibility condition
$(\Phi_x)_t$=$(\Phi_t)_x$ leads to the zero curvature equation
${\cal U}_t- {\cal V}_x+[{\cal U},{\cal V}]=0$ which yields the
integrable generalized coupled NLS equation (\ref{gptwo}).

\begin{figure}
\includegraphics[scale=0.75]{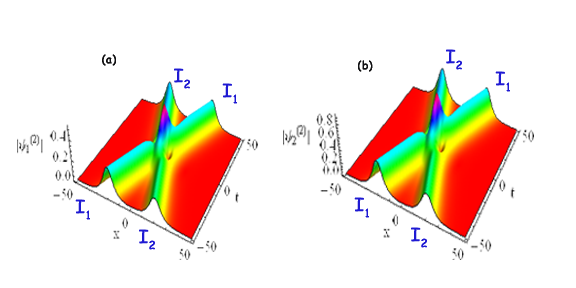}
\caption{Elastic collision of solitons in the celebrated Manakov
model for $a \equiv c =1$ and $b \equiv d = 0$,
$\alpha_{10}=0.15$, $\alpha_{20}=0.15$, $\beta_{10}=0.25$,
$\beta_{20}=0.25$, $\chi_{1}=2$, $\chi_{2}=3$, $\delta_{1}=4$,
$\delta_{2}=5$, $\varepsilon_1^{(1)}=0.5$,
$\varepsilon_1^{(2)}=0.5$ }\label{figtwosol1}
\end{figure}

\begin{figure}
\includegraphics[scale=0.6]{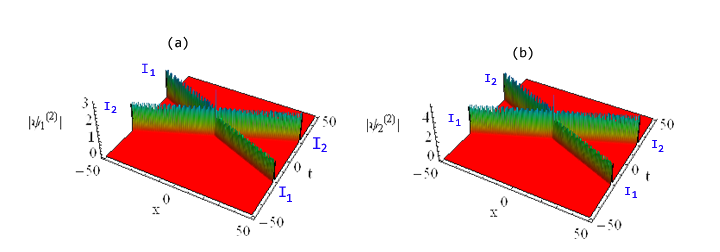}
\caption{FWM induced rotation and enhancement of intensities of
solitons for  $a \equiv c=1$ and $b=d=0.5$ with the other
parameters as in fig.(\ref{figtwosol1})
}\label{figtwosol2}
\end{figure}

\begin{figure}
\includegraphics[scale=0.6]{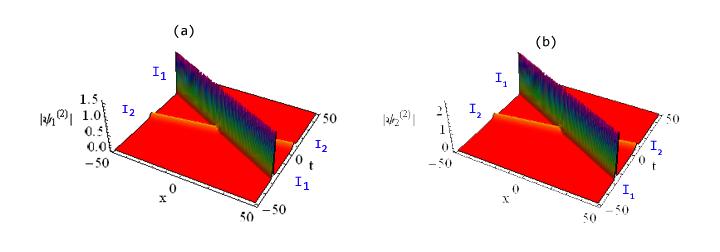}
\caption{Enhancement of the intensity $I_{1}$ by manipulating
$d=3.5$ keeping the other parameters as in fig.(\ref{figtwosol2})
except $\varepsilon_1^{(1)}=0.8$ and $\varepsilon_1^{(2)}=0.5$
}\label{figtwosol3}
\end{figure}

\begin{figure}
\includegraphics[scale=0.6]{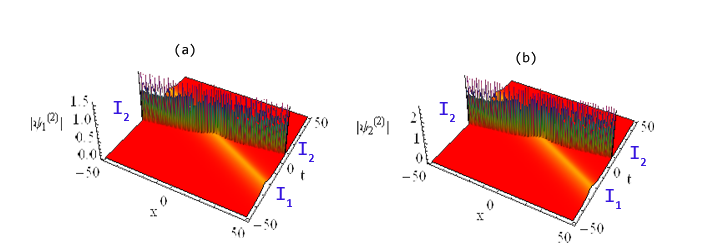}
\caption{Enhancement of intensity $I_{2}$ by manipulating $b=3.5$
keeping other parameters as in fig.(\ref{figtwosol2}) except
$\varepsilon_1^{(1)}=0.5$ and $\varepsilon_1^{(2)}=0.8$
}\label{figtwosol4}
\end{figure}

Recently Agalarov et.al.,\cite{agalarov} investigated the GCNLS
equation.(\ref{gptwo}) for $d = b^{\ast}$ and transformed it to
Manakov model. It should be mentioned that equation.(\ref{gptwo})
with four independent arbitrary parameters $a,c,b$ and $d$ can
also be mapped onto the celebrated Manakov model only for the
parametric choice $b=d$ and $ac-bd=\sigma=\pm1$ under the
following transformation
\begin{equation}
\psi_{1}=q_{1}-d q_{2}, \quad \psi_{2}=a q_{2}.
\end{equation}
so that we obtain equation.(\ref{nlsmode1}) (after suitable
algebraic manipulation) with $g_{11}=g_{21}=a$, $g_{12}=g_{22}=a
\sigma$.

We again emphasize  that GCNLS equation.(\ref{gptwo}) offers the
freedom to choose arbitrary $a,b,c$ and $d$ and the Manakov and
modified Manakov model only arises as a special case of
equation.(\ref{gptwo}). We would like to emphasize that any
conversion of GCNLS equation.(\ref{gptwo}) to either Manakov (or
modified Manakov) model deprives us the freedom to choose $a,b,c$
and $d$ arbitrarily.

\section{Bright Solitons and Collisional Dynamics}
\quad It is worth to pointing out at here that the above type of
equation such as Davey-Stewartson (DS) equation has been
investigated by different methods like first integral method
\cite{Integralmethod}, variational iteration method
\cite{variational} and decomposition method \cite{decomposition},
but gauge transformation approach \cite{llchaw1991} which is
employed to investigate the model equations. (\ref{gptwo}) is more
effective and handy to generate multi soliton solution. Employing
gauge transformation approach one obtains the bright soliton
solution of the following form
\begin{align}
\psi_{1}^{(1)} &= 2 \varepsilon _{1}^{(1)} \beta_{1}
\mathrm{sech}(\theta
_{1})e^{i(-\xi _{1})},  \label{onesol1} \\
\psi_{2}^{(1)} &= 2 \varepsilon _{2}^{(1)} \beta_{1}
\mathrm{sech}(\theta _{1})e^{i(-\xi _{1})}, \label{onesol2}
\end{align}%
where
\begin{align}
\theta _{1} &= 2 x \beta _{1}-4 \int (\alpha _{1}\beta
_{1}) dt + 2 \delta_{1},\label{theta} \\
\xi _{1} &= 2 x \alpha _{1}- 2 \int (\alpha _{1}^{2}-\beta
_{1}^{2}) dt -2\chi _{1}\label{phase}.
\end{align}%
with $~\alpha _{1}=\alpha _{10}(a \tau_1^2 + b \tau_1 \tau_2 + d
\tau_1 \tau_2 +c \tau_2^2)$, $\beta _{1}=\beta _{10}(a \tau_1^2 +
b \tau_1 \tau_2 + d \tau_1 \tau_2 +c \tau_2^2)$  while $\delta
_{1}$, $\chi _{1}$, $\tau_1$  and $\tau_2$ are arbitrary
parameters and $\varepsilon _{1}^{(1)}, \varepsilon _{2}^{(1)}$
are coupling parameters, subject to the constraint $|\varepsilon
_{1}^{(1)}|^{2}+|\varepsilon _{2}^{(1)}|^{2}=1$.

It is obvious from the above  that the amplitude of the bright
solitons depends not only on the SPM and XPM parameters $a$ and
$c$, but also on FWM parameters $b$ and $d$. This freedom in the
system parameters can be manipulated to switch energy between two
light pulses or between two bound states of a given light pulse.
The gauge transformation approach \cite{llchaw1991} can be
extended to generate multisoliton solution. For example, the
two-solition solution $\psi_{1,2}^{(2)}$ for the two modes can be
expressed as
\begin{subequations}
\begin{align}
\psi_{1}^{(2)} = 2 I \frac{A1}{B},\\
\psi_{2}^{(2)} = 2 I \frac{A2}{B},
\end{align}\label{twosolsolution}
\end{subequations}

where
\begin{align}
A1&= M_{121} M_{222} \left(\zeta_2 -\zeta_1 \right) \left(\zeta_1
-\bar{\zeta_1}\right) \left(\zeta_2-\bar{\zeta_2}\right)+M_{122}
M_{221}\left(\zeta_2-\bar{\zeta_1}\right)\left(\bar{\zeta_2}-\zeta_1\right) \left(\zeta_2-\bar{\zeta_2}\right)\notag\\
&+ M_{111} M_{122} \left(\zeta_2-\bar{\zeta_1}\right) \left(\bar{\zeta_2}-\bar{\zeta_1}\right) \left(\zeta_2-\bar{\zeta_2}\right)+M_{112} M_{121} \left(\zeta_1-\bar{\zeta_1}\right) \left(\bar{\zeta_2}-\zeta_1\right) \left(\bar{\zeta_2}-\bar{\zeta_1}\right), \notag\\
A2&= M_{112} M_{211} \left(\zeta_2-\zeta_1\right) \left(\zeta
_1-\bar{\zeta_1}\right) \left(\zeta_2-\bar{\zeta_2}\right)+M_{111}
M_{212} \left(\zeta_2-\bar{\zeta_1}\right)\left(\bar{\zeta_2}-\zeta_1\right) \left(\zeta_2-\bar{\zeta_2}\right)\notag\\
&+ M_{212} M_{221} \left(\zeta_2-\bar{\zeta_1}\right) \left(\bar{\zeta_2}-\bar{\zeta_1}\right) \left(\zeta_2-\bar{\zeta_2}\right)+M_{211} M_{222} \left(\zeta_1-\bar{\zeta_1}\right) \left(\bar{\zeta_2}-\zeta_1\right) \left(\bar{\zeta_2}-\bar{\zeta_1}\right),\notag\\
B&= \left(M_{122} M_{211}+M_{121} M_{212}\right)
\left(\zeta_1-\bar{\zeta_1}\right)\left(\zeta_2-\bar{\zeta_2}\right)+\left(M_{112}
M_{221}+M_{111}M_{222}\right) \left(\zeta_2-\bar{\zeta_1}\right)\notag\\
&\left(\bar{\zeta_2}-\zeta_1\right)
\left(M_{111}M_{112}+M_{221}M_{222}\right)\left(\zeta_2-\zeta_1\right)\left(\bar{\zeta_2}-\bar{\zeta_1}\right),\notag
\end{align}
with  $\zeta_2 = \bar{\zeta_2}^* = \alpha_2 + i \beta_2$,
\begin{align}
M_{11j}&= e^{-\theta_j}\sqrt{2};\quad\nonumber
M_{12j}=e^{-i\xi_j}\varepsilon_1^{(j)};\quad\nonumber
M_{13j}=e^{-i\xi_j}\varepsilon_2^{(j)};\nonumber\\
M_{21j}&= e^{i\xi_j}\varepsilon_1^{*(j)};\quad\nonumber
M_{22j}=e^{\theta_j}/\sqrt{2};\quad\nonumber
M_{23j}=0;\nonumber\\
M_{31j}&= e^{i\xi_j}\varepsilon_2^{*(j)};\quad\nonumber
M_{32j}=0;\quad\nonumber M_{33j}=e^{\theta_j}/\sqrt{2},\nonumber
\end{align}
where
\begin{align}
\theta _{j} &= 2  \beta _{j} x - 4\int (\alpha _{j}\beta_{j})dt+2\delta_{j}, \label{thetaj}\\
\xi _{j} &= 2 \alpha _{j} x-
2\int(\alpha_{j}^{2}-\beta_{j}^{2})dt-2\chi _{j},\label{xij}
\end{align}
and $j = 1, 2$


The two soliton solution given by
equations.(\ref{twosolsolution}-\ref{xij}) can be rewritten
asymptotically (i.e) at $t=\pm \infty$ in the following form
\cite{motopaper}\\

\textbf{Before collision:}
\begin{align}
\psi_{(1)}^{(2-)} &= A_1^{(2-)} \varepsilon _{1}^{(1)} [\frac
{{C_1 + C_2 + C_3 + C_4}}{B_1 + B_2}] \rm{e}^{\left(i \xi_1 \right
)},\notag \\
\psi_{(2)}^{(2-)} &= A_2^{(2-)} \varepsilon _{1}^{(1)} [\frac
{{C_1 + C_2 + C_3 + C_4}}{B_1 + B_2}] \rm{e}^{\left(i \xi_1
\right)}, \notag\label{beforecolli} \\
\end{align}

\textbf{After collision:}
\begin{align}
\psi_{(1)}^{(2+)}&= A_1^{(2+)} \varepsilon _{1}^{(2)} [\frac {{C_1
+ C_2 + C_3 + C_4}}{B_1 + B_2}] \rm{e}^{\left(i (\xi_2-\xi_1)
\right)},\notag \\
\psi_{(2)}^{(2+)}&= A_2^{(2+)} \varepsilon _{1}^{(2)} [\frac {{C_1
+ C_2 + C_3 + C_4}}{B_1 + B_2}] \rm{e}^{\left(i (\xi_2-\xi_1)
\right)}, \label{aftercolli}\notag\\
\end{align}
In the above expression, the (-) and (+) sign indicates before and
after collision and the subscript and superscript depicts the
component (mode) and soliton respectively. with,
\begin{align}
A_1^{(2-)} &= \alpha_1 .\Big[
\frac{\zeta_1-\zeta_2^{\ast}}{\zeta_1-\zeta_2}\Big],\quad
A_2^{(2-)} = \beta_1 .\Big[
\frac{\zeta_1-\zeta_2^{\ast}}{\zeta_1-\zeta_2}\Big],\notag\\
A_1^{(2+)} &= \alpha_2.\Big[
\frac{\zeta_2-\zeta_1^{\ast}}{\zeta_2-\zeta_1}\Big],\quad
A_2^{(2+)} =\beta_2.\Big[
\frac{\zeta_2-\zeta_1^{\ast}}{\zeta_2-\zeta_1}\Big].
\end{align}

where
\begin{align}
C_1&= \{-2\beta _2 [(\alpha_2 - \alpha_1 )^2  - (\beta_1^2  -
\beta_2^2 )]- 4i\beta_1 \beta_2 (\alpha_2  - \alpha_1 )\}
\rm{e}^{(\theta_1^{\ast}+\theta_1  + i\xi_2 )},\nonumber\\
C_2&= - 2\beta_2 [(\alpha_2  - \alpha_1 )^2  + (\beta_1^2  +
\beta_2^2 )]\rm{e}^{(\theta_1 - \theta_1^{\ast}  + i\xi_2 )},\nonumber\\
C_3&= \{- 2\beta_1 [(\alpha_2  - \alpha_1 )^2  + (\beta_1^2  -
\beta_2^2 )] + 4i\beta_1 \beta_2 (\alpha_2  - \alpha_1 )\}
\rm{e}^{(i\xi_1  + \theta_2^{\ast}+\theta_2 )}, \nonumber \\
C_4&= -4i\beta_1 \beta_2 [(\alpha_2  - \alpha_1 ) - i(\beta_1  -
\beta_2 )]\rm{e}^{(i\xi_1 +\theta_2 - \theta_2^{\ast} )},\nonumber\\
B_1&= -4\beta _1 \beta _2 [\sinh (\chi_1 )\sinh (\chi_2 ) +
\cos (\xi _1  - \xi _2)],\nonumber\\
B_2&=2\;\cosh (\chi_1 )\;\cosh (\chi_2 )\;[(\alpha _2  - \alpha _1
)^2  + (\beta _1^2  + \beta _2^2 )], \nonumber
\end{align}

with $~\alpha _{j}=\alpha _{j0}(a \tau_1^2 + b \tau_1 \tau_2 + d
\tau_1 \tau_2 +c \tau_2^2)$, $\beta _{j}=\beta _{j0}(a \tau_1^2 +
b \tau_1 \tau_2 + d \tau_1 \tau_2 +c \tau_2^2)$ and the notation
$\theta_j-\theta_j^{\ast}=2 i \beta_j x - 4 (\alpha_j^2-
\beta_j^2)t$ and $\theta_j^{\ast}+\theta_j= 2 \beta_j x- 4
\alpha_j \beta_j t$  where $j=1,2$.

If we choose
\begin{equation}
\frac{\alpha_{10}}{\alpha_{20}}=\frac{\beta_{10}}{\beta_{20}},
\label{elasticasymptcondition}
\end{equation}
with $\varepsilon _{1}^{(1)}=\varepsilon _{1}^{(2)}$ (or)
$\varepsilon _{1}^{(1)}\neq \varepsilon _{1}^{(2)}$ keeping the
condition $|\varepsilon _{1}^{(j)}|^2+|\varepsilon
_{2}^{(j)}|^2=1$, $j=1,2$, one observes elastic collision of
bright solitons. Any violation of  the above condition given by
equation.(\ref{elasticasymptcondition}) results in inelastic
collision of solitons leading to the exchange of energy between
two optical beams (modes).

\begin{figure}
\includegraphics[scale=0.6]{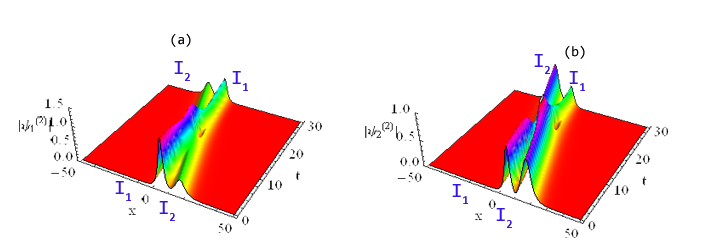}
\caption{Inelastic collision of the solitons in the Manakov model
for $a \equiv c=1$ and $b\equiv d = 0$ with $\alpha_{10}=0.1$,
$\alpha_{20}=0.25$, $\beta_{10}=0.2$, $\beta_{20}=0.3$,
$\chi_{1}=0.2$, $\chi_{2}=0.3$, $\delta_{1}=0.1$,
$\delta_{2}=0.2$, $\varepsilon_1^{(1)}=0.5$,
$\varepsilon_1^{(2)}=0.8$ }\label{figie1}
\end{figure}

\subsection{Intramodal collision of bright solitons}

We first consider the elastic collision of solitons in the Manakov
model shown in figure.(\ref{figtwosol1}) under the parametric
choice given by equation.(\ref{elasticasymptcondition}). The
corresponding amplitudes of two soliton solution before and after
collision can be rewritten as

\begin{align}
A_j^{(2-)} &= \left(%
\begin{array}{c}
\alpha_1 \\
\beta_1 \\
\end{array}%
\right)(a \tau_1^2 +c \tau_2^2)) \varepsilon _{1}^{(1)} .
\frac{\zeta_1-\zeta_2^{\ast}}{\zeta_1-\zeta_2},\notag\\
A_j^{(2+)} &=\left(%
\begin{array}{c}
\alpha_1 \\
\beta_1 \\
\end{array}%
\right)(a \tau_1^2 +c \tau_2^2))\varepsilon _{1}^{(2)}.
\frac{\zeta_2-\zeta_1^{\ast}}{\zeta_2-\zeta_1}, \quad
j=1,2 \notag\\
\end{align}
When we introduce FWM, the amplitudes of two solitons solution
asymptotically take the  following form

\begin{align}
A_j^{(2-)} &= \left(%
\begin{array}{c}
\alpha_1 \\
\beta_1 \\
\end{array}%
\right)(a \tau_1^2 +(b+d)\tau_1 \tau_2 +c \tau_2^2) \varepsilon
_{1}^{(1)} .
\frac{\zeta_1-\zeta_2^{\ast}}{\zeta_1-\zeta_2},\notag\\
A_j^{(2+)} &= \left(%
\begin{array}{c}
\alpha_1 \\
\beta_1 \\
\end{array}%
\right)(a \tau_1^2 +(b+d)\tau_1 \tau_2  +c \tau_2^2)\varepsilon
_{1}^{(2)}. \frac{\zeta_2-\zeta_1^{\ast}}{\zeta_2-\zeta_1},\quad
j= 1,2
\end{align}
From the above, it is obvious that  the introduction of FWM
parameters $b$ and $d$ contributes to the enhancement of
intensities as shown in figure.\ref{figtwosol2}. In addition, one
observes the rotation of the trajectories of solitons.

Now, to enhance the intensity of a given bound state in a given
mode (optical pulse), we manipulate the FWM parameter say $d$ and
choose unequal coupling parameter $\varepsilon _{1}^{(1)}=0.8$ \&
$\varepsilon _{1}^{(2)}=0.5$. The corresponding density profile
shown in figure.(\ref{figtwosol3}) enhances the intensity of one
bound state $(I_1)$ at the expense of the other $(I_2)$ in each
mode (optical pulse). By manipulating the real parameter $b$
instead of $d$ and keeping $\varepsilon _{1}^{(1)}=0.5$ \&
$\varepsilon _{1}^{(2)}=0.8$, the density profile is shown in
figure.(\ref{figtwosol4}), where one observes the enhancement of
$I_2$ at the expense of $I_1$ in each mode (optical beam).

\begin{figure}
\includegraphics[scale=0.6]{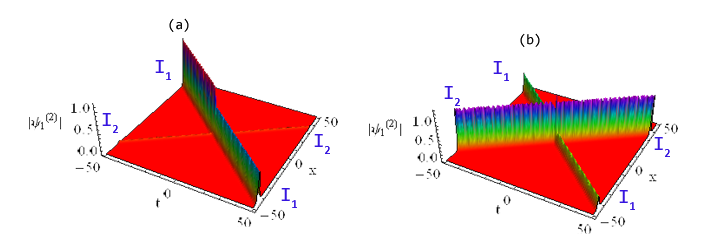}
\caption{Rotation of bright solitons and switching of energy for
$b=0.5$ and $d=3.5$ keeping the other parameters the same as in
fig.\ref{figie1}except $\varepsilon_1^{(1)}=0.8$ and
$\varepsilon_1^{(2)}=0.5$}\label{figie4}
\end{figure}

\begin{figure}
\includegraphics[scale=0.6]{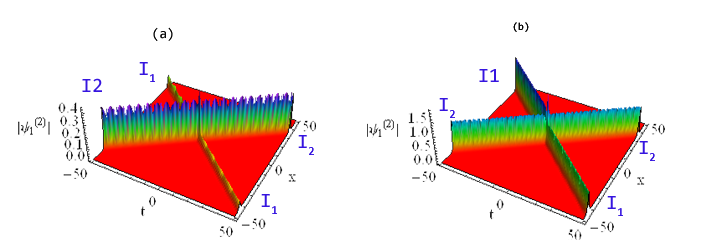}
\caption{Reversal of intensity enhancement for $b=3.5$ and $d=0.5$
keeping the other parameters the same as in
fig.\ref{figie1}}\label{figie5}
\end{figure}

\subsection{Intermodal collision of bright solitons}
To manipulate the intensity of a given optical beam (mode), we now
begin with the celebrated inelastic collision of solitons in the
Manakov model \cite{kanna} shown in figure.(\ref{figie1}) in the
absence of FWM. The asymptotic form of two solitions can be
written as

\begin{align}
A_j^{(2-)} &=\left(%
\begin{array}{c}
\alpha_1 \\
\beta_1 \\
\end{array}%
\right)(a \tau_1^2 +c \tau_2^2)) \varepsilon _{1}^{(1)} .
\frac{\zeta_1-\zeta_2^{\ast}}{\zeta_1-\zeta_2},\notag\\
A_j^{(2+)} &=\left(%
\begin{array}{c}
\alpha_2 \\
\beta_2 \\
\end{array}
\right)(a \tau_1^2 +c \tau_2^2))\varepsilon _{1}^{(2)}.
\frac{\zeta_2-\zeta_1^{\ast}}{\zeta_2-\zeta_1},\notag\\
\end{align}
When we introduce unequal FWM now and reverse the choice of
coupling parameters (i.e.,) $\varepsilon _{1}^{(1)}=0.8$ \&
$\varepsilon _{1}^{(2)}=0.5$, the intensity redistribution shown
in figure.(\ref{figie4}) shows that one can manipulate energy in a
given bound state of the beam desirably.

Interchanging the values of the FWM \& coupling parameters results
in the reversal of intensity distribution of the solitons as shown
in fig.\ref{figie5}.

The above results indicate that one can not only manoeuvre the
intensity distribution between the light beams, but also
manipulate the intensity distribution of the given bound state in
a given mode (optical pulse) and this freedom arises due to a
suitable combination of FWM, SPM and XPM. It is also worth
pointing out that the introduction of unequal SPM and XPM
alongwith FWM in the above interaction of solitons contributes to
a marginal increase of intensities besides rotating the
trajectories of solitons. From the above, we also observe that the
intensity redistribution among the bound states of a given optical
beam (mode) or between two optical beams (modes) through
manipulation of FWM parameters $b$ \& $d$ is always accompanied by
rotation to sustain the stability of solitons.

When we neglect SPM and XPM in equation.(\ref{gptwo}) (a=c=0), we
obtain the following coupled NLS equation
\begin{align}
i \psi_{1t}+\psi_{1xx}+2( b \psi_1 \psi_2^{\ast} + d \psi_2 \psi_1^*)\psi_1=0, \nonumber\\
i \psi_{2t}+\psi_{2xx}+2( b \psi_1 \psi_2^{\ast} + d \psi_2
\psi_1^*)\psi_2=0. \label{gptwonew}
\end{align}
It should be mentioned that equation.(\ref{gptwonew}) arises only
as a special case of equation.(\ref{gptwo}) and the phenomenon of
soliton reflection and noninteraction of solitons \cite{motopaper}
can also be obtained as a special case. It should also be
mentioned that under suitable transformation, the model governed
by GCNLS equation.(\ref{gptwo}) can be mapped onto its counterpart
in Gross-Pitaevskii (GP) equation which means that one can switch
matter wave intensities desirably and this mechanism can  be
employed for matter wave switching.

\section{Discussion}
In this paper, we have derived a generalized coupled  NLS (CGNLS)
equation containing four arbitrary real parameters with two real
parameters corresponding to SPM and XPM and the other two real
parameters accounting for FWM. The collisional dynamics of bright
solitons shows that one can have the luxury of sustaining
desirable intensity in a given bound state of the optical beam or
in a given optical pulse and the celebrated Manakov does not have
the same freedom. It should be emphasized that the manipulation of
light intensities in the above GCNLS equation (\ref{gptwo})
explicitly depends on the interaction of light with the  medium
unlike the Manakov model where the intensity redistribution occurs
by changing the parameters associated with the phase of solitons.
Our investigation may open the floodgates for optical and matter
wave switching in nonlinear optics and BECs. It would be
interesting to study the ramifications of complex FWM parameters
on the dynamics of bright solitons.

\section{Acknowledgements}

Authors would like to acknowledge Dr. Telman Gadzhimuradov in
sharing his perspective in improving the contents of the paper.
PSV  wishes to thank Department of Science and Technology (DST)
for the financial support. RR wishes to acknowledge the financial
assistance received from DST (Ref.No:SR /S2/HEP-26/2012), UGC
(Ref.No:F.No 40-420/2011(SR), Department of Atomic Energy
-National Board for Higher Mathematics (DAE-NBHM) (Ref.No: NBHM /
R.P.16/2014/Fresh dated 22.10.2014) and Council of Scientific and
Industrial Research (CSIR) (Ref.No: No.03(1323)/14/EMR-II dated
03.11.2014) dated 4.July.2011) for the financial support in the
form Major Research Projects. KP thanks the DST, NBHM, IFCPAR,
DST-FCT and CSIR, Government of India, for the financial support
through major projects.



\begin{thebibliography}{30}

\bibitem{nlseqn} Hasegawa A,  Tappert F. Transmission of
stationary nonlinear optical pulses in dispersive dielectric
fibers. I. Anomalous dispersion. Appl. Phys. Lett 1973; 23: 142.

\bibitem{nlseqn2} Hasegawa A, Tappert F. Transmission of stationary
nonlinear optical pulses in dispersive dielectric fibers. II.
Normal dispersion. Appl. Phys. Lett  1973; 23: 171.

\bibitem{fibre} Agrawal G P. Nonlinear Fiber Optics. San Diego: Academic
Press; 2001.



\bibitem{fibre2} Hasegawa A,  Kodama Y. Solitons in Optical
Communication. New York: Oxford University Press; 1995.



\bibitem{painleve} Sahadevan R, Tamizhmani K M, Lakshmanan M. Painleve analysis
and integrability of coupled non-linear Schrodinger equations. J.
Phys. A: Math.Gen 1986; 19: 1783.

\bibitem{manakov}Manakov S V. On the theory of two dimensional stationary self-focusing of electromagnetic waves. Zh. Eksp. Teor. Fiz 1973;
65: 505-516.




\bibitem{manakov2}Kaup D J, Malomed B A. Soliton trapping and daughter waves
in the Manakov model. Phys. Rev. A  1993; 48: 599.





\bibitem{modifymanakov}
Makhankov  V G, Makhaldiani  N V, Pashaev  O K. On the
integrability and isotopic structure of the one-dimensional
Hubbard model in the long wave approximation. Phys. Lett.A  1981;
 81: 161.

\bibitem{modifymanakov2}Tsoy E N, Akhmediev  N. Dynamics and interaction of pulses
in the modified Manakov model. Opt. Commun. 2006; 266: 660.

\bibitem{rkrishnan}
Radhakrishnan R, Lakshmanan M, Hietarinta J. Inelastic collision
and switching of coupled bright solitons in optical fibers. Phys.
Rev. E  1997; 56: 2213.


\bibitem{psv:pre:nls}
Radha R, Vinayagam P S, Porsezian K. Rotation of the trajectories
of bright solitons  and realignment of intensity distribution in
the coupled nonlinear schr\"{o}dinger equation. Phys. Rev. E 2013;
88: 032903.

\bibitem{FR}
Malomed B A. Soliton Management in Periodic Systems. New York:
Springer; 2006.

\bibitem{review}
Radha R, Vinayagam P S. An analytical window into the world of
ultracold atoms. Rom. Rep. Phys. 2015; 67 (1): 89.

\bibitem{park.shin}Park Q-H, Shin H J. Painlev\'{e} analysis of the coupled nonlinear Schr\"{o}dinger
equation for polarized optical waves in an isotropic medium. Phys.
Rev. E  1999; 59: 2737.

\bibitem{motopaper} Wang D-S, Zhang D-J,  Yang J.
Integrable properties of the general coupled nonlinear Schrödinger
equations. J. Math. Phys 2010; 51: 023510.

\bibitem{agalarov} Agalarov A, Zhulego V,
Gadzhimuradov T. Bright, dark, and mixed vector soliton solutions
of the general coupled nonlinear Schr$\rm\ddot{o}$dinger
equations. Phys. Rev. E 2015; 91: 042909.

\bibitem{Integralmethod}
Jafari H, Sooraki A, Talebi Y, Biswas A. The first integral method
and traveling wave solutions to Davey–Stewartson equation.
Nonlinear Analysis: Modelling and Control. 2012; 17 (2):182.

\bibitem{variational}
Jafari H, Kadem A, Baleanu D, Yilmaz T. Solution of the fractional
Davey- Stewartson equations with variational iteration method.
Rom. Rep Physics. 2012; 64 (2): 337.

\bibitem{decomposition}
Jafari H, Tajadodi  H, Bolandtalat A, Johnston S J. A
Decomposition Method for Solving the Fractional Davey-Stewartson
Equations. Int. J. Appl. Comput. Math. (2015); 1: 559.

\bibitem{llchaw1991}
Chau L -L , Shaw J C, Yen H C. An alternative explicit
construction of N-soliton solutions in 1+1 dimensions. J. Math.
Phys. 1991; 32: 1737.




\bibitem{kanna}
Malomed B A. Inelastic collisions of polarized solitons in a
birefringent optical fiber. J. Opt. Soc. Am. B  1992;  9: 2075.


\end{thebibliography}
\end{document}